# SIMULATION STUDIES OF THE BACKSCATTERING SIGNAL IN HSRL TECHNIQUE

**Angelika Georgakopoulou**

*NTUA,Dept of Physics,NTUA,Zografou Campus,157 80 Athens Greece, agelika@central.ntua.gr*

**ABSTRACT**

The technique of High Spectral Resolution Lidar (HSRL) for atmospheric monitoring allows the determination of the aerosol to molecular ratio and can be used in UHECR Observatories using air fluorescence telescopes. By this technique a more accurate estimate of the Cherenkov radiation superimposed to the fluorescence signal can be achieved. A laboratory setup was developed to determine the backscattering coefficients using microparticles diluted in water and diffusion interfaces. In this setup we used a CW SLM laser at 532 nm and a 250 mm Newtonian telescope.

Simulations of the above experimental configuration have been made using Scatlab$^©$, FINESSE$^©$ 0.99.8 and MATLAB$^©$ and are presented in this work. We compare the simulated 2-dimensional Fabry-Perot fringe images of the backscattering signal recorded in the CCD sensor with that of experimental ones. Additionally, we simulated the backscattering of the laser beam by the atmosphere at a height of 2000 m and we have studied the influence of the beam and its diameter on the fringe image.

## 1. INTRODUCTION

The atmospheric monitoring by the HSRL Technique in UHECR Observatories can be used to determine the aerosol to molecular ratio in the atmosphere allowing more accurate estimate of the Cherenkov radiation superimposed to the fluorescence signal. Comparison studies of the backscattering signal in the detector using MATLAB$^©$, Scatlab$^©$ and FINESSE 0.99.8$^©$ simulation programs with the experimental one, are given in this work.

-MATLAB$^©$ is the well known numerical computation program, owned by The Matheworks,Inc. Here we mostly make use of the incorporated Image Processing Toolbox.

-Scatlab$^©$ is a freely distributed Mie and T-matrix scattering simulation tool, owned by V. Bazhan [1]. It makes use of classical Mie theory to simulate light scattering by spherical (or not spherical, even coated) particles. The user enters values for the several parameters involved and Scatlab returns several scattering plots, polarization/depolarization data, and values for the scattering, extinction and backscattering cross sections.

- Finesse$^©$ 0.99.8 is a freely distributed interferometer simulation program, owned by Andreas Freise [2]. The user can create a virtual interferometer with several components (such as lasers, spaces, mirrors etc) and simulate the results of altering one or more parameters. The data are returned in the form of MATLAB executable files, from which the user can draw results and create plots. The mathematical viewpoint behind Finesse lies in viewing the laser beam as traveling wave of the form

$$E=E_0\exp(i(\varphi+\omega t))=A\exp(i\omega t)$$

where t is time, $\omega$ is the angular frequency and A denotes the beam's complex amplitude. At the start and the end of each component, the mutual coupling between amplitudes is given by linear equations. These form a linear system that is solved with the help of the SPARSE package [4]. The characteristics of the outgoing beam are computed this way.

## 2. MICROPARTICLE SCATTERING SIMULATION

The backscattering coefficient is obtained by a scattering simulation with Scatlab$^©$. The resulting backscattering coefficient is found to be $\beta_{sim}= 0.65096*10^{-6}$ m$^{-1}$sr$^{-1}$, which is very close to the experimental value $\beta_{exp}= 0.64334*10^{-6}$ m$^{-1}$sr$^{-1}$ we find in the following section. The backscattered beam enters the etalon and we simulate the result in Finesse$^©$. The following figures demonstrate the characteristics of the beam and are comparable to the experimental results displayed in Section 3.

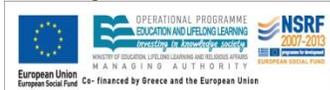

**This research has been co-financed by the European Union (European Social Fund – ESF) and Greek national funds through the Operational Program "Education and Lifelong Learning" of the National Strategic Reference Framework (NSRF) - Research Funding Program: Heracleitus II. Investing in knowledge society through the European Social Fund.**

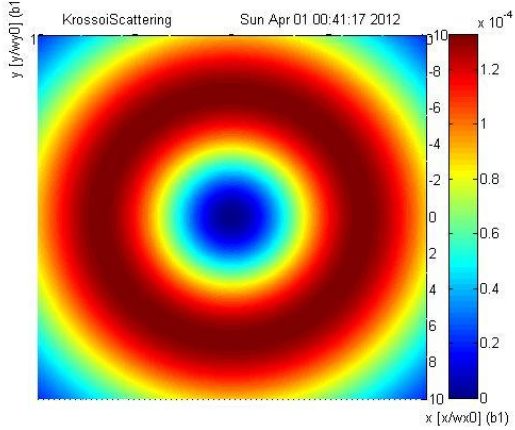

Figure 1. Interference fringes for the simulated scattered beam. Comparable with Figure 4.

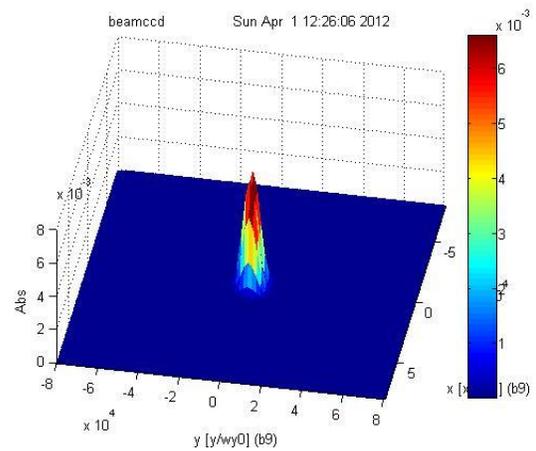

Figure 3. A rough intensity plot of the simulated backscattered beam. Comparable with Figure 6.

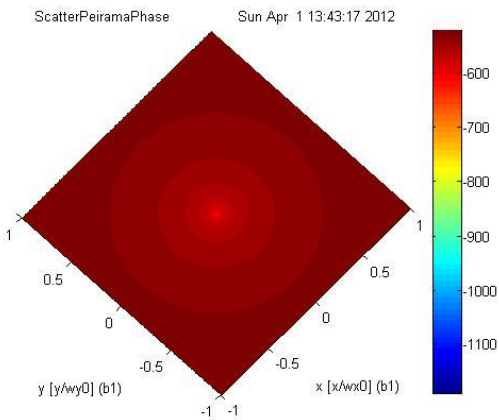

Figure 2. A plot of the scattered beam's amplitude in dB.

## 3. EXPERIMENTAL SCATTERING RESULTS

A DPSSL CW SLM laser beam at 532 nm is scattered through a volume of water containing spherical microparticles of size 1 μm. The beam backscatters into a detecting array that includes of a Newtonian telescope (250 mm), a Fabry-Perot etalon (50 mm) and a CCD [3]. The CCD photos were analysed with the graphical program FV (Figure 4). Also, using MATLAB's Image Processing Toolbox, a larger plot of the interference fringes was created, along with a plot of the beam's intensity profile (Figure 6). We find that Figure 4 and Figure 1 are comparable, as are Figure 6 and Figure 3.

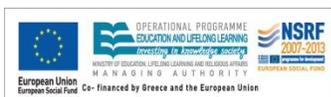

**This research has been co-financed by the European Union (European Social Fund – ESF) and Greek national funds through the Operational Program "Education and Lifelong Learning" of the National Strategic Reference Framework (NSRF) - Research Funding Program: Heracleitus II. Investing in knowledge society through the European Social Fund.**

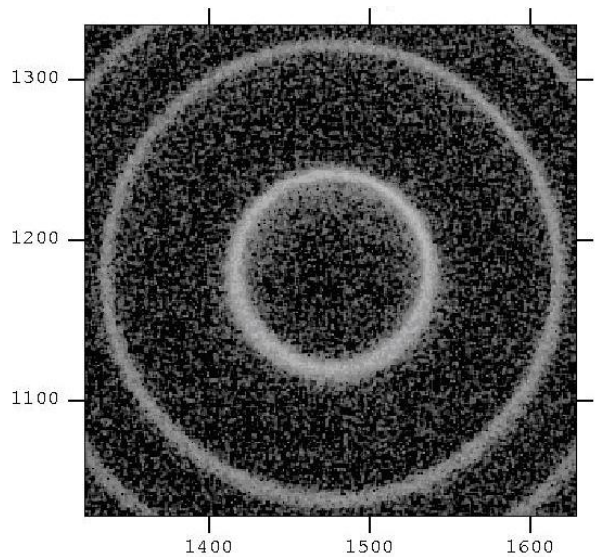

Figure 4. Interference fringes for the CCD photo obtained with the program FV. We observe the similarity between these and the interference fringes obtained via simulation (Figure 1).

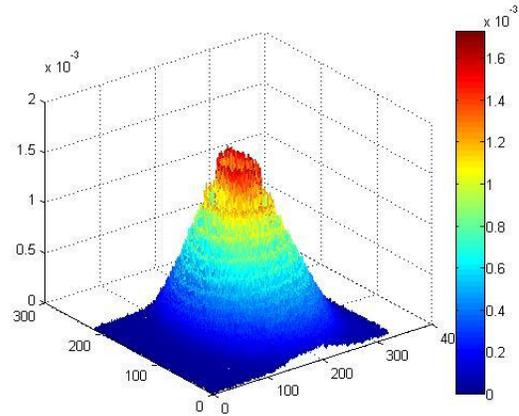

Figure 6. The intensity plot of the CCD photo, obtained with MATLAB's Image Processing Toolbox.

## 4. ATMOSPHERIC BACKSCATTERING SIMULATION

A hypothetical Gaussian laser beam is scattered in the atmosphere at 2000 m. At that height, the extinction coefficient is equal to $\beta=1.1892*10^{-6}$ $m^{-1}sr^{-1}$ (PapayannisData@2000m). Knowing this, we simulate the bistatic setup in [3] with Finesse$^{©}$ and provide here some results:

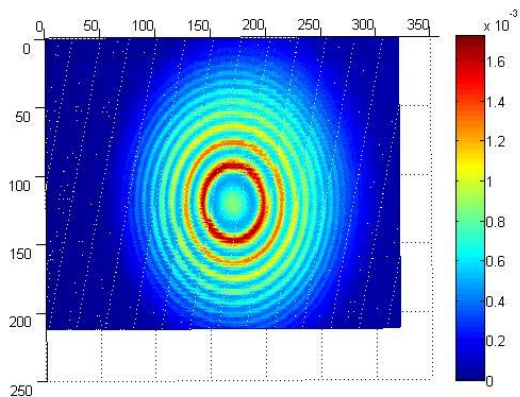

Figure 5. Interference fringes obtained by analysing the CCD photo with the help of MATLAB's Image Processing Toolbox.

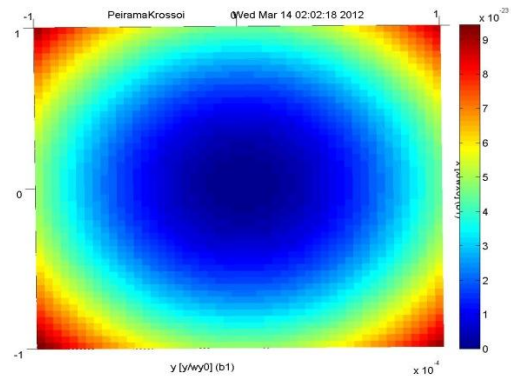

Figure 7. Interference fringes for the scattered beam. With a transmitting power of 0.1 W, the beam's power is of order of magnitude equal to $10^{-13}$, hence the rough plot. See Figure 8.

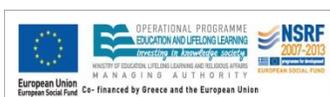

**This research has been co-financed by the European Union (European Social Fund – ESF) and Greek national funds through the Operational Program "Education and Lifelong Learning" of the National Strategic Reference Framework (NSRF) - Research Funding Program: Heracleitus II. Investing in knowledge society through the European Social Fund.**

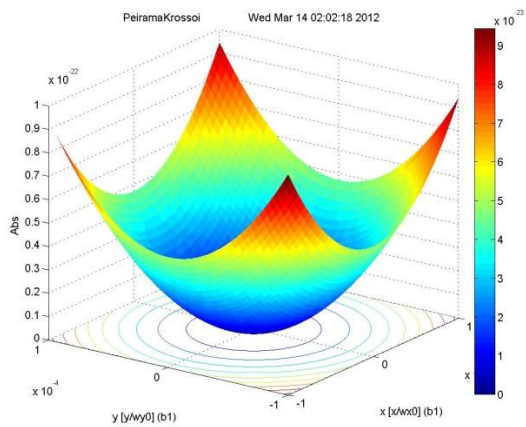

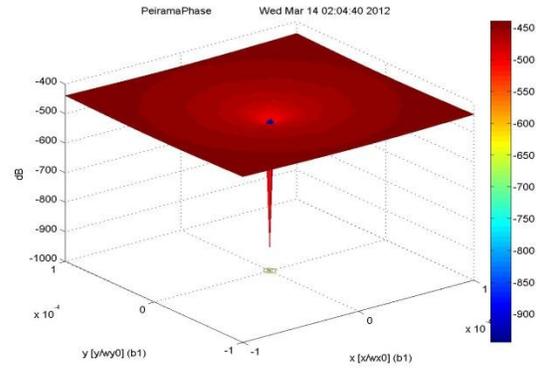

Figure 10. The scattered beam's amplitude in dB from a different viepoint, obtained with MATLAB.

Figure 8. A plot of the scattered beam's amplitude. The level projections are the interference fringes of Figure 7.

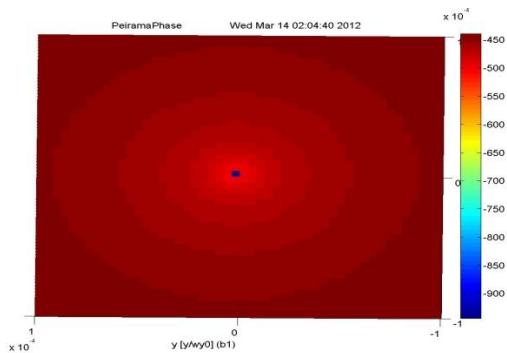

Figure 9. A plot of the scattered beam's amplitude in dB.

## CONCLUSIONS

A laboratory scattering test of the High Spectral Resolution LIDAR method was conducted. The same setup was simulated with the help of Scatlab$^©$ and Finesse 0.99.8$^©$. We find that the simulated results are comparable to experimental data that involve intensity plots, interference fringes and the backscattering coefficient. Also, the results of a second simulation involving HSRL scattering in the atmosphere are presented. However, there is still more work to be done, computational as well as laboratorial.

## REFERENCES

1. URL: www.scatlab.org

2. A. Freise: 'FINESSE, Frequency domain interferometer simulation software (1999-2010)'. (www.gwoptics.org/finesse/)

3. E. Fokitis, S. Maltezos, A. Papayannis, P. Fetfatzis, A. Georgakopoulou, and A. Aravantinos, "High Spectral Resolution LIDAR Receivers to measure Aerosol to Molecular Scattering Ratio in Bistatic mode for use in Atmospheric Monitoring for EAS Detectors", Nuclear Physics B (Proc. Suppl.) 197 (2009) 317-321.

4. K.S.Kundert, A. Sangiovanni-Vincentelli: "SPARSE: A sparse linear equation solver.", University Of California, Berkeley (1988). 3,116.

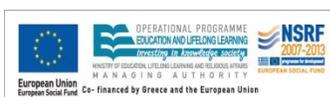

**This research has been co-financed by the European Union (European Social Fund – ESF) and Greek national funds through the Operational Program "Education and Lifelong Learning" of the National Strategic Reference Framework (NSRF) - Research Funding Program: Heracleitus II. Investing in knowledge society through the European Social Fund.**